\documentclass[useAMS,usenatbib]{mn2e}

\usepackage{amsmath,amssymb,amsfonts,bm}
\usepackage{graphicx}
\usepackage{hyperref}
\usepackage{calc}
\usepackage{color}
\usepackage{wrapfig}
\usepackage{url}
\usepackage{paralist}
\usepackage{array}
\usepackage{tabularx}
\usepackage{txfonts}
\title[Offset between DM and OM in galaxy clusters]{Offset between dark matter and ordinary matter: evidence from a sample of 38 lensing clusters of galaxies}
\author[HuanYuan Shan et al.]{HuanYuan Shan$^{1,6}$\thanks{E-mail:
shanhuany@gmail.com, qinbo@bao.ac.cn}, Bo Qin$^{1 \star}$, Bernard
Fort$^{2}$, Charling Tao$^{3}$, Xiang-Ping Wu$^{1}$
\newauthor and HongSheng Zhao$^{1,4,5}$\\
$^{1}$National Astronomical Observatories, Chinese Academy of Sciences, Beijing 100012, China\\
$^{2}$UPMC Universit\'e Paris 06, UMR7095, Institut dAstrophysique
de Paris, F-75014, Paris, France\\
$^{3}$Centre de Physique des Particules de Marseille,
CNRS/IN2P3-Luminy and Universit\'e de la M\'editerran\'ee, Case 907,
F-13288 Marseille Cedex 9,
France\\
$^{4}$SUPA, University of St Andrews, KY16 9SS, UK\\
$^{5}$Leiden University, Sterrewacht and Instituut-Lorentz, Niels-
Bohrweg 2, 2333 CA, Leiden, The
Netherlands\\
$^{6}$Department of Astronomy, School of Physics, Peking University,
Beijing, 100871, China }

\begin{document}

\date{Accepted \dots . Received \dots; in original form \dots}


\maketitle

\label{firstpage}

\begin{abstract}
We compile a sample of 38 galaxy clusters which have both X-ray and
strong lensing observations, and study for each cluster the
projected offset between the dominant component of baryonic matter
center (measured by X-rays) and the gravitational center (measured
by strong lensing). Among the total sample, $45\%$ clusters have
offsets $>10''$. The $>10''$ separations are significant,
considering the arcsecond precision in the measurement of the
lensing/X-ray centers. This suggests that it might be a common
phenomenon in unrelaxed galaxy clusters that gravitational field is
separated spatially from the dominant component of baryonic matter.
It also has consequences for lensing models of unrelaxed clusters
since the gas mass distribution may differ from the dark matter
distribution and give perturbations to the modeling. Such offsets
can be used as a statistical tool for comparison with the results of
$\Lambda$CDM simulations and to test the modified dynamics.
\end{abstract}

\begin{keywords}
dark matter-gravitational lensing-X-rays: galaxies: clusters
\end{keywords}

\section{Introduction}

Seventy years after Zwicky's first piece of evidence for dark matter
(DM) in galaxy clusters, the physics model for DM still remains
ambiguous, ranging from Weakly-Interacting Massive Particles (WIMPs)
to gravity-modifying Tensor-Vector-Scalar fields (TeVeS, Bekenstein
2004).  In-between these seemingly conflicting theories, we also
have models where the DM changes properties in different
environments due to gravitational polarization or interactions with
a dark energy field (Blanchet \& Le Tiec 2009, Li \& Zhao 2009).
These in-between DM models explain why DM in galaxies seems to
satisfy the MOND formulae of Milgrom (1983) with a common empirical
scale $a_0 \sim \sqrt{\Lambda} \sim 10^{-10}{\rm m}\,{\rm s}^{-2}$
found by fitting galaxy rotation curves.  They are also consistent
with the cosmic microwave background, and the late time dark energy
effect $\Lambda$ or order $a_0^2$.

MOND has gained enormous momentum, partly for its success in making
reasonable stable galaxies and explaining galactic phenomenology
(e.g., Wang et al. 2008; Wu et al. 2009; Gentile et al. 2009). MOND
can also give high velocity encounters of galaxy clusters (Llinares
et al. 2009). However, it does not fully account for the discrepancy
between the X-ray and dynamical mass in rich clusters of galaxies
(Gerbal et al. 1992; The \& White 1988; Aguirre, Schaye \& Quataert
2001; Sanders 2003; Tian, Hoekstra, \& Zhao 2009), which Sanders
(2003) explained by introducing a 2~eV neutrino component. We also
should note that the environmental-dependent DM of Li \& Zhao (2009)
and the DM dipoles of Blanchet \& Le Tiec (2009), both mimic MOND
for galaxies, might resolve the apparent contradiction of MOND for
clusters.

A big challenge to modified gravity is the observations of the
bullet cluster 1E0657-56 (Bradac et al. 2006; Clowe et al. 2006;
Markevitch et al. 2006). Weak lensing observations of the bullet
cluster, combined with earlier X-ray measurements, clearly indicated
that the gravitational field of the cluster has an obvious offset
from its ordinary matter distribution.

One immediate question one may ask is: Is the bullet cluster the
only system that uniquely shows the spatial separation between dark
and ordinary matter? Is the phenomenon of DM-baryon separation so
rare in the universe, or could it be more common?

In galaxy clusters, most baryons (or ordinary matter) exist in the
form of diffuse X-ray emitting gas. The stellar component is larger
at the cluster center where bright galaxies are concentrated but DM
is still the dominant component. This has been demonstrated by
Gavazzi et al. (2003) and Gavazzi (2005): for the inner $\lesssim
100$~kpc regions of a lensing cluster, the stellar component only
occupies a few percent of the total mass. Therefore, the X-ray
images could be used as a reasonable approximation of the ordinary
matter distribution in a cluster. X-ray observations measure
directly the ordinary matter distributions, while the total
projected mass distributions (mainly DM) can be measured by
gravitational lensing. Thus, a comparison between X-ray and lensing
observations of galaxy clusters potentials may unveil possible differences
between dark and ordinary matter distributions, just as the
observations of the bullet cluster have revealed.

Strong lensing has the potential to determine the cluster mass
center with arcsecond precision. The $0.5''$ high spatial resolution
of the Chandra X-ray satellite means that we may determine an
accurate position of a cluster's baryonic center, though the X-ray
data processing may eventually give rise to a larger uncertainty of
no more than a few arcseconds (see e.g. Smith et al. 2005). Thus, we
can compare the gravitational and baryonic centers of galaxy
clusters by investigating a fairly large sample. If the
gravitational center of a lensing cluster does not match the
ordinary matter center, we could say that a separate DM component
might exist. Indeed we expect this result for cluster on-going major merger.

Offsets between lensing and X-ray centers were incidently
noticed a decade ago by Allen (1998) when he was studying a sample
of 13 clusters. However, no attempt has been made to use such
offsets as a dynamical signature, which might be used as
quantitative measure of the quality of the DM model by comparing
with similar statistics coming from $\Lambda$CDM simulations.

In this paper, we compile a sample of 38 galaxy clusters that have
both strong lensing (SL) and X-ray observations. We carefully
check the lensing hypothesis and location of the main potential of
the lens if several deflectors are considered. Combining the lensing
data with X-ray data, we obtain for each cluster an offset between
the lensing center and X-ray center on the projected plane. We use
this offset as an ``indicator'' to describe the dark matter-ordinary
matter separation. Our data strongly support the idea that the
gravitational potential in clusters is mainly due to a non-baryonic
fluid, and any exotic field in gravitational theory must resemble
that of CDM fields very closely. Moreover, we find that unrelaxed
clusters have larger offsets than relaxed clusters.

Interestingly, simulations of CDM+baryon for galaxy clusters in the
standard $\Lambda$CDM cosmology has recently found offsets between
the baryonic and DM centers in clusters (Forero-Romero et al. 2010).
The offset can be as large as 100~kpc. This roughly supports our
findings, though the details of the cluster distribution functions
between the simulations and our observational data show some
difference.

\section{Cluster Sample and Results}

The sample of $38$ clusters is listed in Table~1, which shows the
positions of the lensing/optical and X-ray centroids on the
projected plane, the two-dimensional offsets between the lensing and
X-ray centers (given in~arcsecond), and the corresponding angular
diameter distances (in~kpc) calculated in the $\Lambda$CDM
cosmology. The lensing/optical data are from the HST WFPC2 archive,
the SDSS survey and some other independent observations. Most of the
X-ray data are from the Chandra archive, except for 13 clusters from
ROSAT (marked with ``9'',``11'',``15'' in the Ref.B column). For
clusters with several DM clumps, we choose the one with the largest
X-ray temperature and give the corresponding X-ray center in the
sample. The clusters in our table are classified from their X-ray
morphologies as relaxed clusters (with cooling flows) and unrelaxed
clusters (which are dynamically unmature). This definition has been
used in the literature by Allen (1999), Wu (2000), Baldi et al
(2007), and Dunn \& Fabian (2008). Figure~1 shows the cluster number
counts as a function of the offset (2-D separation between the
lensing and X-ray centers).

\begin{figure}
\centering
\includegraphics[width=1.0\columnwidth]{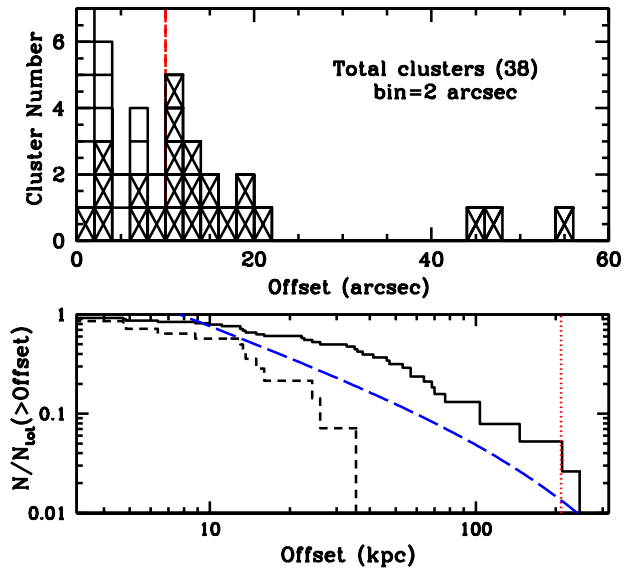}
\caption{\rm {\em Upper panel}: Histogram of the cluster
distribution as a function of the offset between the lensing center
and X-ray center, for a sample of 38 galaxy clusters with both
strong lensing and X-ray observations. The crossed boxes represent
unrelaxed clusters. The red vertical dashed line marks the offset of
10'.  {\em Lower panel}: Cumulative distribution of the total
clusters (solid line) and relaxed clusters (dashed line) as a
function of the 2-D physical separation between lensing and X-ray
centers. The red vertical dotted line corresponds to the bullet
cluster offset. The blue long dashed line
$P_{2D}=0.04 \left(\frac{d}{200}\right)^{-1}\exp\left(\frac{d}{200}\right)$ shows the Forero-Romero et al.
(2010) simulation results.} \protect\label{fig.p}
\end{figure}

The lensing centers of these clusters are always determined by their
arcs and images' positions. Remarkably about half of the best models
for relaxed clusters have adopted the ``Brightest Cluster  Galaxy''
(BCG) center of mass as the DM halo center. The BCG has a faint
extended star halo with a distribution similar to that of the DM
(same center and almost the same ellipticity and main axis
orientation). For the case where the DM center is actually kept as a
free parameter, the offset between the BCG and lensing center seems
to be below $1.5''$. It is reasonable to consider the BCG center as
the lensing center of relaxed clusters. As for the unrelaxed
clusters which show larger offsets, the BCGs cannot be taken as a
unique lensing center. Smith et al. (2005) show that the offset
errors of the main potential of $7$ unrelaxed clusters are with the
typical value around $2''$, which includes uncertainties on the
central coordinates of the cluster mass distribution in the relevant
lens model with all the arcs and images' positions.

The X-ray centers of the clusters are always determined by the
weighted averaging of the pixel centers covering the cluster, which
depends on the number of X rays in a pixel, the number of pixels and
flux distribution used for the centroid determination. Lazzati \&
Chincarni (1998) have estimated the effect of pixel sizes on the
centroid determined to be around a quarter of a pixel. For pixel
sizes equivalent to $6''$, the effect is less than $2''$.  The error
from the algorithm fit on the centroid can however be larger, but
the authors claim that a value of $5''$ is pretty conservative.

The X-ray center determination may be also dependent on the size of
the region considered for the estimates, especially for unrelaxed
clusters whose X-ray morphologies are not so regular compared with
relaxed clusters. To test this, Maughan et al. (2008) have measured
the ``centroid shifts'' for 115 X-ray clusters, following earlier
work by Mohr et al. (1993), O'Hara et al. (2006) and Poole et al.
(2006). The centroid shifts were determined in a series of circular
apertures centered on the cluster X-ray peak, with the radius of the
apertures decreasing in steps of $5\%$ from $R_{500}$ to
$0.05R_{500}$. Maughan et al. (2008) found that the value of the
centroid shift, $w$, ranges from a few tenths to a few of
$10^{-2}R_{500}$ in the sample. Typically $R_{500}\sim 1$~Mpc for
clusters. So this centroid shift is roughly equivalent to measuring
the difference between the very extreme cases of the $r\sim 0$
region and the $r=1$~Mpc region. For the clusters in our sample, $w$
is typically a few arcsecs. For example, the unrelaxed clusters A68
and A209 have $w = 4''$ and $2''$ respectively. For the well-known
unrelaxed bullet cluster, $w = 8''$.

Considering the above errors as a whole, we define our {\it
Selection Criterion---We only regard offsets larger than $10''$ as
being significant}. This is a fairly conservative criterion if we
consider lensing rays tracing with simulations like the Millennium
(Hilbert et al. 2007). For systems with offsets much smaller than
$10''$, observational uncertainties may start to play an important
role as well as the lensing modeling errors, if external shear
perturbations have not been well identified in the field.

One prominent feature of Figure~1 is that all the clusters with
offsets $>10''$ are unrelaxed clusters, suggesting that the
mechanism for this lensing/X-ray offset is probably related to
dynamical relaxations, which is the same as the research about the
offset between BCG/X-ray center (Sanderson et al. 2009). In total,
there is a rather high percentage of clusters with obvious offsets
--- almost half of the clusters ($45\%$) in the whole sample have
offsets $>10''$, and 3 clusters (or $8\%$) even have extraordinarily
large separations of $>40''$, indicating that such offsets may not
be rare in unrelaxed clusters. Taking into account the fact that the
observed offsets are only two-dimensional, the true separations
(3-D) should be even higher (see the discussion section). In
addition to the well-studied bullet cluster, which has an offset
value of $47.4''$, clusters A2163, and A2744 also have very large
offsets of $44.0''$ and $54.3''$, respectively.

Our selection criterion of $>10''$ is mainly based on the errors in
the lensing and X-ray measurements, which are typically about a few
arcsecs. If we increase this threshold from $10''$ to $20''$, then
our sub-sample of large offset clusters will decrease drastically
from $45\%$ to only $10\%$. This is naturally expected, because the
distribution (total number) of clusters decreases very rapidly with
increasing offset. This trend is clearly demonstrated by the blue
dashed curve in Figure~1 which is a result of numerical simulations.
Also, dynamical relaxations will tend to reduce the offset in
clusters, leading to more relaxed clusters. Interestingly, all of
the four offset~$>20''$ clusters are discovered at relatively high
redshifts of $z>0.2$.

We also note some peculiar cases in our sample. A370 seems made of
two relaxed clusters along the same line of sight which have not yet
merged. In such a case, despite the blending effect, we see two
X-ray peaks at each BCG center (equivalent to lensing centers, c.f.
the map of Kneib et al. 1993). In A370, the distance between the two
equivalent potential wells is $40''$ (according to the lens
modeling). Bonamente et al. (2006) showed a map of the blend
emissions of two cluster lenses. A detailed study of the offset
might involve subdividing the sample according to different degrees
of merging. Further studies of the ordinary matter offsets might
involve subdividing the cluster sample according to different
degrees of merging by considering the velocity distributions of
galaxy members. A1682 contains two BCG, and only one of them
produces strong lensing arc. The small radius of the known arc (Sand
et al. 2005) indicates that this is galaxy-scale lensing, and is not
caused primarily by the dark matter core of the cluster. So it would
be incorrect to use the SL center as the mass center of this
cluster. The unrelaxed cluster A1914 only has one arc without
redshift and counter-arc (Sand et al. 2005), which would result in
large errors in SL center determination. The offset between SL and
X-ray center is $53.6''$. Dahle et al. (2002) studied this cluster
with weak lensing, and showed that the cluster has a triangle of
bright elliptical (with probably two cD) galaxies. The weak lensing
measurement places the center of mass of the cluster very close to
the X-ray centroid, indicating that the SL is caused by one of the
cD in a complex merging process involving several substructures and
possible projection effects. For such a cluster the weak lensing
appears representative of the cluster center on large scales. But
without other multiple gravitational arcs around cDs it is not
possible to conclude about the DM peaks at the cluster center. Only
the BCG-X ray offset gives the signature of a merging process.

\section{Discussion and Conclusions}
\label{sect:discussion}

The mass in stars of a cluster is small compared to its X-ray gas.
Therefore the X-ray images can be used as a reasonable approximation
of ordinary matter distribution in a cluster. Similarly the stellar
mass component is not dominant in SL modeling even if it often seems
the SL and stellar mass centers peak at the same place, the BCG
center. Consequently we have used the X-ray images (which is the
result of the diffuse intracluster X-ray gas) to find the center of
ordinary matter distribution and the lensing mass to find the center
of dark matter distribution. Indeed this is an approximation and we
have explained that the error in the separation angle shall be much
less than $10''$. Therefore the separation between DM and ordinary
matter is highly significant for the unrelaxed clusters.

It is noticeable that all the clusters in our sample are clusters
with $z>0.1$ (most of them have $z>0.2$). This selection effect is
caused by strong lensing clusters because higher redshift clusters
have higher lensing probabilities. In order to have high precision
determinations of the gravitational center of clusters, we have to
focus on strong lensing clusters---which means that our cluster
sample has to be a high redshift sample instead of containing many
local clusters. In principle, we should have a more unbiased sample
with sufficient low redshift clusters which are measured by, e.g.,
high quality weak lensing.

However, current weak lensing determination of the lensing center is
much less robust compared with strong lensing. Indeed, we have
investigated clusters with both X-ray and weak lensing observations
(but not strong lensing) and found the lensing/X-ray offsets. For
example, cluster MS1054 (Jee et al. 2005) has an offset of $19.5''$.
MS1008 (Ettori \& Lombardi 2003) has an offset of $5.43''$. But
unfortunately, the errors from weak lensing are much larger.
Nevertheless, high quality weak lensing observations of clusters,
especially low redshift clusters, will be of particular interest.

It should be noted that the offset here is only 2-D, i.e., the
separation on the projected plane. The true separation (3-D) could
be much larger. One extreme example is cluster CL0024+17. The
redshift distribution of the cluster's member galaxies revealed that
the configuration of this cluster is along the line of sight (Czoske
et al. 2001). Moreover, recent studies suggested that this cluster
may have undergone a head-on collision along the line of sight (Jee
et al. 2007; Qin et al. 2008). So its 3-D separation is probably
much larger than the 2-D offset in Table~1. The other extreme case
is the bullet cluster, where the configuration (as well as the
head-on collision) is roughly on the projected plane, indicating
that its 3-D separation is close to the 2-D offset.

The bullet cluster has provided us a system that clearly shows the
existence of a DM component. RX J1347.5-1145 (Bradac et al. 2008) is
similar but with another line of sight projection: the east massive
clump has lost all its X-ray gas. For the bullet cluster, Angus et
al. (2007) have pointed out that MOND could be rescued if DM is made
of 2 eV neutrinos, following the $\mu$HDM model introduced by
Sanders (2003). Indeed such an observation seems to match structure
formation in both $\mu$HDM and $\Lambda$CDM cosmologies. Meanwhile,
Knebe et al. (2009) found that MOND can in principle produce offsets
of effective DM, but the offsets are often small, about 1~kpc.

There are also implications in the CDM framework. The significant
offset found here is a signal of the merging process and is a
measure of the departure from equilibrium, and consequently the
X-ray determined dynamical mass based on equilibrium would
under-predict compared to lensing-determined mass, which does not
use assumptions of equilibrium (Allen 1998; Smith et al. 2005; Zu
Hone et al. 2009).

Concerning the SL modeling, it is necessary to identify all the mass
distribution that might induce an external shear on the arc system.
But it might be also important to include the shear-like
perturbation of an offset gas component, a possible systematic
effect neglected in previous models. All SL modeling (except the
bullet cluster) has been continuously done with a smooth halo
component which includes DM component plus the gas with the implicit
hypothesis that the gas follows the DM distribution, which is not
true except for a fully relaxed cluster. Most sophisticated models
also considered the mass perturbation of the stellar component and
subhalo associated to early type galaxy members (sometimes with a
separate stellar component of the BCG) to improve the modeling of
the arc configuration (Kneib et al. 1996; Meneghetti et al. 2003;
Keeton 2003; Limousin et al. 2007; Natarajan et al. 2007). Despite
that Allen (1998) has explicitly noted the offset of the ordinary
gas matter, it is remarkable that nobody has considered the SL
modeling with an offset of a large proportion of the ordinary mass,
on the same footing as the member galaxies effect. In this paper, we
strongly argue that the offset of the ordinary gas matter, which
represents $10\%-20\%$ of the total cluster mass, should be figured
out explicitly for unrelaxed clusters.

In summary, our finding of a high percentage ($45\%$) of clusters
with offsets $>10''$ in the whole sample suggests that it might be a
common phenomenon that in unrelaxed galaxy clusters the
gravitational field is more or less separated from the ordinary
matter distribution. Such separation is best explained if a
non-baryonic matter component (DM) does exist. The separations are
probably due to dynamical relaxations, as all the clusters with
offsets $>10''$ in the sample are unrelaxed clusters.

Indeed simulations of the cluster gas in the CDM framework do show
the existence of the large offsets (Ferero-Romero et al. 2010). The
simulation (Gottloeber \& Yepes 2007), called The Marenostrum
Universe, was run using the code GADGET2 and followed the adiabatic
evolution of gas and dark matter from $z=40$ to $z=0$ in a comoving
cube of $500 \rm h^{-1}$~Mpc. The simulation predicts a median
offset of about 18~kpc, which is in general agreement with our whole
sample. Nevertheless, the profiles of these offset distributions
(i.e., cluster distribution functions with respect to offset) are
nonidentical, as shown in Figure~1. A K-S test shows that the
significance of differences are $99.2\%$ and $99.6\%$ for the whole
sample and the subsample of relaxed clusters, respectively.

From Figure~1, the biggest difference between the simulation results
and our total cluster sample is that our sample has more clusters
with large offsets. The main reasons could be as follows:

(1) Our sample of lensing clusters is biased towards higher redshift
while the simulations are not. Obviously, dynamical relaxation will
reduce the offset in clusters, leading to fewer clusters with large
offsets as compared with the high redshift sample. In other words,
most clusters in our sample have $z>0.2$, and it is possible that
these high redshift clusters are not fully evolved dynamically as
compared with low redshift clusters. Also our sample of 38 clusters
is still not big enough, and a larger sample is needed to draw a
more robust conclusion.

(2) When a cluster of a given total mass is not fully relaxed and
has still large merging clumps about to merge at the center, the
length of the caustic lines is increased as compared to a fully
relaxed cluster. So it could result in a higher probability to
produce lensing arcs, as has been pointed out by Meneghetti et al.
(2007). Therefore the fact that we observe more distant clusters
which are less relaxed than at $z=0$ has a double bias effect.

(3) The gas physics in galaxy clusters is complicated and less
well-understood. Obviously, different treatment of the gas could
result in different cluster mass profiles and baryonic
distributions, which could give different offset values.

There are other possible explanations, e.g., the merging history
model may need to be revised, or the simulations may not have the
complete recipes for the physics of DM. Another example is that DM
could be possibly coupled to a dark energy scalar field, which
distorts the non-linear dynamics at the centers of halos without
ruining the large scale success of the standard cold DM. In short,
more comparisons of new simulations and larger samples of clusters
would be very rewarding in the future, as the offsets observed here
provide astrophysicists a new quantitative tool to evaluate
different cosmologies.

\chapter{\flushright{\bf{Acknowledgments}}}
\flushleft{We thank Raphael Gavazzi, Zhi-Ying Huo, Marceau Limousin
and Shude Mao for discussions, and an anonymous referee for helpful
suggestions. HYS and BQ are grateful to the CPPM for hospitality.
HSZ acknowledges partial support from the Dutch NWO visitor's grant
\#040.11.089 to Henk Hoekstra. This work was supported by the
National Basic Research Program of China (973 Program) under grant
No. 2009CB24901, and CAS grants KJCX3-SYW-N2 and KJCX2-YW-N32.}

\onecolumn

\begin{table}
\centering \caption{\rm Offsets between the lensing centroids and
X-ray centroids of a sample of 38 clusters. The lensing centroid
data are taken from the HST WFPC2 archive, SDSS survey and some
other independent measurements of lensing clusters. The X-ray
centroid data are taken from the Chandra archive (except for 13
clusters denoted by ``9'',``11'',``15'' in Ref.B that are taken from
the ROSAT archive). Ref.A and Ref.B give the references of the
lensing and X-ray data respectively. The last column shows the
classification of the clusters: ``R/U'' means relaxed/unrelaxed.}
\begin{tabular}{lllrclrcrrc}
\hline
Cluster & $z_{\rm cluster}$ &\multicolumn{2}{c}{Lensing/optical centroid} & Ref.A & \multicolumn{2}{c}{X-ray centroid} & Ref.B &\multicolumn{2}{c}{Offset} & Class\\
\hline
   &     & \multicolumn{1}{c}{R.A.} & \multicolumn{1}{c}{Dec.}  & &
  \multicolumn{1}{c}{R.A.}  & \multicolumn{1}{c}{Dec.}  &  & (arcsec) & (kpc) & \\
\hline
1E0657-56 & 0.296 & 06 58 35.3 & -55 56 56.3 & 1 & 06 58 30.2 & -55 56 35.9 & 1&47.4 & 209.2& U\\
A68 & 0.255 & 00 37 06.9 & +09 09 23.3 & 3 & 00 37 06.2 & +09 09 33.2 &12& 14.3 & 56.7 & U\\
A209 & 0.206 & 01 31 52.5 & -13 36 40.5 & 3,4 & 01 31 53.5 & -13 36 46.1 &13& 15.6 & 52.7 & U\\
A267 & 0.230 & 01 52 42.0 & +01 00 26.2 & 3& 01 52 42.1 & +01 00 35.7 &12& 9.62 & 35.3 & U\\
A370 & 0.375 & 02 39 53.1 & -01 34 54.8 & 2,5& 02 39 53.2 & -01 34 35.0 &12& 19.9 & 102.7 & U\\
A383 & 0.187 & 02 48 03.4 & -03 31 45.2 & 3& 02 48 03.4 & -03 31 46.2 &13& 1.0 & 3.13 & R\\
A586 & 0.170 & 07 32 20.3 & +31 38 00.1 & 2& 07 32 20.2 & +31 37 55.6 &12& 4.68 & 13.6 & R\\
A697 & 0.282 & 08 42 57.6 & +36 21 59.1 & 2&08 42 57.8 & +36 21 57.2 &13& 3.07 & 13.1 & U\\
A773 & 0.217 & 09 17 53.4 & +51 43 37.2  & 3& 09 17 52.8 & +51 43 40.4 &13& 6.43 & 22.6 & U\\
A963 & 0.206 & 10 17 03.6 & +39 02 49.2 & 3&10 17 03.7 & +39 02 56.2 &14& 7.10 & 24.0 & R\\
A1682 & 0.234 & 13 06 49.9 & +46 33 33.5 & 2, 17, 18&13 06 51.1 & +46 33 29.5 &13& 12.6 & 46.9 & U\\
A1689 & 0.183 & 13 11 29.5 & -01 20 27.6 & 6&13 11 29.5 & -01 20 28.2 &12& 0.60 & 1.85 & U\\
A1763 & 0.228 & 13 35 20.1 & +41 00 04.0 & 3&13 35 20.0 & +40 59 53.8 &14& 10.3 & 37.6 & U\\
A1835 & 0.252 & 14 01 02.1 & +02 52 42.3 & 3&14 01 02.0 & +02 52 41.7 &12& 1.61 & 6.33 & R\\
A1914 & 0.171 & 14 26 01.0 & +37 49 45.0 & 2, 20&14 26 01.2 & +37 49 34.0 &13& 11.3 & 32.9 & U\\
A2204 & 0.151 & 16 32 46.9 & +05 34 33.1 & 2&16 32 46.9 & +05 34 31.9 &12& 1.20 & 3.15 & R\\
A2163 & 0.203 & 16 15 49.1 & -06 08 43.0 & 15, 19&16 15 46.2 & -06 08 51.3 &12& 44.0 & 146.9 & U\\
A2218 & 0.176 & 16 35 49.3 & +66 12 45.3 & 3&16 35 51.9 & +66 12 34.5 &12& 19.1 & 56.9 & U\\
A2219 & 0.228 & 16 40 19.8 & +46 42 41.7 & 3&16 40 20.2 & +46 42 31.2 &14& 11.3 & 41.2 & U\\
A2259 & 0.164 & 17 20 09.7  & +27 40 07.4 & 2&17 20 08.5 & +27 40 11.0 &12& 16.3 & 45.9 & U \\
A2261 & 0.224 & 17 22 27.2 & +32 07 57.5 & 2&17 22 27.1 & +32 07 57.8 &12& 1.31 & 4.72 & R\\
A2294 & 0.178 & 17 24 12.6 & +85 53 11.6 & 2&17 24 09.6 & +85 53 11.0 &13& 3.28 & 9.87 & U\\
A2390 & 0.228 & 21 53 36.9 & +17 41 43.4 &2& 21 53 36.5 & +17 41 45.2 &14& 6.00 & 21.9 & U\\
A2744 & 0.308 & 00 14 20.8 & -30 24 03.0 &15& 00 14 18.7 & -30 23 16.0 &16& 54.3 & 246.3 & U\\
AC114 & 0.313 & 22 58 48.3 & -34 48 07.2 & 2,7 & 22 58 49.2 & -34 48 27.0 &14&22.7 & 104.1 & U\\
CL0024 & 0.395 & 00 26 35.0 & +17 09 43.0&8 & 00 26 35.9 & +17 09 40.0 &14& 13.2 & 70.4 & U\\ 
MS0016 & 0.541 & 00 18 33.6 & +16 26 16.0 &2& 00 18 33.6 & +16 26 14.6 &14& 1.40 & 8.90 & R\\
MS0440 & 0.190 & 04 43 09.7 & +02 10 19.5 &2& 04 43 09.8 & +02 10 19.5 &9& 1.50 & 4.89 & R\\
MS0451 & 0.550 & 04 54 10.6 & -03 00 50.7 &2& 04 54 11.4 & -03 00 52.7 &12& 12.1 & 76.8 & U\\
MS1137 & 0.782 & 11 40 22.3 & +66 08 14.1 &2& 11 40 22.3 & +66 08 16.1 &12& 2.00 & 14.9 & R\\
MS1358 & 0.329 & 13 59 50.5 & +62 31 06.8 &2& 35 59 50.6 & +62 31 04.1 &12& 2.79 & 13.2 & R\\
MS1455 & 0.258 & 14 57 15.1 & +22 20 34.9 &2,10& 14 57 15.0 & +22 20 37.3 &14& 2.77 & 11.1 & U\\
MS1621 & 0.426 & 16 23 35.2 & +26 34 28.2 &2& 16 23 35.7 & +26 34 19.0 &14& 11.4 & 63.6 & U \\
MS2053 & 0.580 & 20 56 21.4 & -04 37 50.9 &2& 20 56 20.7 & -04 37 51.6 &14& 10.5 & 69.1 & U\\
MS2137 & 0.313 & 21 40 14.9 & -23 39 39.5 &2,10& 21 40 15.2 & -23 39 41.0 &14& 5.70 & 26.1 & R \\
PKS0745 & 0.103 & 07 47 31.3 & -19 17 40.0 &2& 07 47 31.4 & -19 17 46.2 &14& 6.82 & 12.9 & R\\
RXJ1347 & 0.451 & 13 47 30.7 & -11 45 11.0 &15& 13 47 30.6 & -11 45 08.6 &12& 2.81 & 16.2 & R\\
RBS864 & 0.291 & 10 23 39.3 & +04 11 17.0 &11& 10 23 39.6 & +04 11 10.0 &11& 8.32 & 36.3 & R\\
\hline \hline
\smallskip
\end{tabular}
\parbox {6.9in}
{Notes:}
\parbox {6.9in}
{$^a$ Units of R.A. are hours, minutes, and seconds, and units of
Dec. are degrees, arcminutes, and arcseconds, respectively.}
\parbox {6.9in}
{$^b$ References: (1) Clowe et al. 2006; Bradac et al. 2006; (2)
Sand et al. 2005; (3) Smith et al. 2005; (4) Paulin-Henriksson et
al. 2007; (5) Kneib et al. 1993; (6) Limousin et al. 2007; (7)
Campusano et al. 2001; (8) Jee et al. 2007; (9) Gioia et al. 1998;
(10) Newbury \& Fahlman 1999; (11) Kausch et al. 2004; (12)
Bonamente et al. 2006; (13) Maughan et al. 2008; (14) Ota \& Mitsuda
2004; (15) Allen 1998; (16) Kempner \& David 2004; (17) Morrison et
al. 2003; (18) Dahle et al. 2002; (19) Radovich et al. 2008}
\end{table}
\normalsize


\end{document}